\documentclass[5p]{elsarticle}
\usepackage{graphicx}
\begin{document}

\title{\bf New Experiments with Antiprotons
%Grants or other notes
%about the article that should go on the front page should be
%placed here. General acknowledgments should be placed at the end of the article.}
}
%\subtitle{Do you have a subtitle?\\ If so, write it here}

%\titlerunning{Short form of title}        % if too long for running head

%\author{Daniel M. Kaplan %         \and
%        Second Author %etc.
%}

%\authorrunning{Short form of author list} % if too long for running head

\author[First]{Daniel M. Kaplan}
\ead{kaplan@iit.edu}
\address[First]{Physics Division, Illinois Institute of Technology, Chicago, Illinois 60616 USA\\[0.1in]
}

\date{Received: date / Accepted: date}
% The correct dates will be entered by the editor

\begin{abstract}
Fermilab operates the world's most intense antiproton source. Newly proposed experiments can use those antiprotons either parasitically during Tevatron Collider running or after the Tevatron Collider finishes in about 2011. For example, the annihilation of 8 GeV antiprotons might make the world's most intense source of tagged $D^0$ mesons, and thus the best near-term opportunity to study charm mixing and, via {\em CP} violation, to search for new physics. Other potential measurements include sensitive studies of hyperons and of the mysterious $X$, $Y$, and $Z$ states. Production of antihydrogen in flight can be used for first searches for  antihydrogen {\em CPT} violation.
With antiproton deceleration to low energy, an experiment using a Penning trap and an atom interferometer could make the world's first measurement of the gravitational force on antimatter.\end{abstract}

%Include keywords, PACS and mathematical
%subject classification numbers as needed.

\maketitle

\section{Introduction}
\label{intro}
A number of intriguing questions\,---\,many  involving symmetry, the theme of this Workshop\,---\,can be elucidated by a medium-energy antiproton-beam fixed-target experiment. Among these are the possible contributions of physics beyond the standard model to charm mixing and decay, hyperon decay, and the mechanism(s) underlying the $X$, $Y$, and $Z$ states discovered in recent years at the $B$ factories.

 Table~\ref{tab:sens-comp} summarizes the parameters of  current and future  antiproton sources. It can be seen that by far, the highest-energy and highest-intensity antiproton source is at Fermilab. Having formerly served medium-energy antiproton fixed-target experiments, including the charmonium experiments E760 and E835, it is now dedicated entirely to 
 the Tevatron Collider, but could become available again for dedicated antiproton experiments when the Tevatron shuts down (towards the end of 2011 by present estimates).  The CERN Antiproton Decelerator (AD) provides low-energy antiproton beams 
at a tiny fraction of the intensity now available at Fermilab.
Germany's $\approx$billion-Euro plan for the Facility for Antiproton and Ion Research (FAIR) at Darmstadt  includes construction\,---\,yet to be started\,---\,of 30 and 90\,GeV rapid-cycling synchrotrons and low- and medium-energy antiproton storage rings~\cite{FAIR}.
Antiproton operation at  FAIR is anticipated on or after 2017.

\begin{table}\begin{center}\footnotesize
\caption{Antiproton energies and intensities at existing and future facilities.}\label{tab:sens-comp}%\vspace{0.1in}
\begin{tabular}{lccccc}
\hline
\hline
&  & \multicolumn{2}{c}{{Stacking:}} & \multicolumn{2}{c}{{Operation:}}  \\
$\!\!$Facility &${\overline{p}}$ K.E.& {Rate}  & Duty & {Hours} & ${\overline{p}}$/{Yr}\\
 & (GeV) & $(10^{10}/$hr) &  Factor & {/Yr} &  ${(10^{13})}$\\
\hline\hline
& 0.005 \\
\raisebox{1.5ex}[0pt]{$\!\!$CERN AD}& 0.047 & \raisebox{1.5ex}[0pt]{--} & \raisebox{1.5ex}[0pt]{--} & \raisebox{1.5ex}[0pt]{3800} &\raisebox{1.5ex}[0pt]{0.4} \\
\multicolumn{3}{l}{$\!\!$Fermilab Accumulator:}\\
~now& 8 & 20 & 90\% & 5550 & 100\\
~proposed &$\approx$\,3.5--8& 20 & 15\% & 5550 &17 \\
~with new ring &2--20? & 20 & 90\%& 5550 & 100 \\
$\!\!$FAIR ($\stackrel{>}{_\sim}\,$2017) & 2--15 &  3.5 & 90\% & 2780$^*$ & 9\\
\hline\hline
\end{tabular}
\end{center}
\footnotesize
~~$^*$ The lower number of operating hours at FAIR compared with that at other facilities arises from medium-energy antiproton operation having to share time with other programs.
%\vspace{-.25in}
\end{table}

\section{Physics Overview}
In the absence of knowledge about the nature of the sought-for new physics, it is difficult to rank the physics topics mentioned above by  impact and importance. But we can be confident that, should new physics be {\em discovered} in any one of them, it will immediately become the most interesting topic of the day. By current ``handicapping,"  charm mixing is probably highest in priority. The key question is whether there is  new physics in charm mixing, the signature for which is {\em CP} violation~\cite{Bigi-Uraltsev-Petrov}. Despite much effort in the $B$ and $K$ sectors, evidence for physics beyond the standard model has proved elusive. A proper regard for the importance of these issues should  prompt us to look elsewhere as well. 

As pointed out by many authors~\cite{Bigi09,Buchalla-etal}, charm is an excellent venue for such investigation. Not only is  it the only up-type quark for which such effects are possible, but standard-model backgrounds to new physics are suppressed in charm: the CKM factors are small, and the most massive quark participating in loop diagrams is the $b$. As detailed below,  a charm experiment at the Fermilab Antiproton Source might be the world's most sensitive. This is because $\overline{p}p$ or $\overline{p}N$ collisions have an enormous charm-production advantage relative to $e^+e^-$ colliders: charm hadroproduction cross sections are typically $\sim$\,$\mu$b, vs.\  1\,nb for $e^+e^-$.  Of course, luminosity favors $e^+e^-$  (by $\sim 10^2$), and typically  backgrounds do as well. Moreover,   hadroproduction at high energy  has the advantage of longer decay distances. But the countervailing disadvantage is higher multiplicity ($\langle n_{ch}\rangle\sim10$) in the underlying event, which is responsible for the dominant, combinatoric background, suppressed via vertex cuts. The much lower charged-particle multiplicity  ($\langle n_{ch}\rangle\approx2$)  in $\overline{p}p$ collisions  near  open-charm threshold should enable charm samples with cleanliness comparable to that at the $B$ factories, with the application of only modest cuts, and hence, high efficiency. (More details of this argument are presented in Sec.~\ref{sec:charm}.) The competition for this program is LHC$b$ and a possible ``super-$B$ factory,"  which may have significant systematic biases, due e.g.\   to large rates of $b\to c$ decays.

Probably next in priority, the $X(3872)$ has  been observed by several groups (see Table~\ref{tab:X3872}), and is incontrovertibly a real state~\cite{ELQ}. Despite its proximity in mass to various charmonium levels, it does not appear to be one itself. As we will see, $\overline p$$p$ annihilation has the potential to make uniquely incisive measurements of its properties and thereby reveal its true nature.
By scanning the Antiproton Accumulator beam energy across the resonance, Fermilab experiments E760 and E835 made the world's most precise measurements of charmonium masses and widths~\cite{E760-chi,E835-psi-prime}. 
Besides this precision, the other key advantage of the antiproton-annihilation technique is its ability to produce charmonium states of all quantum numbers, in contrast to $e^+e^-$ machines which produce primarily $1^{--}$ states and the few states that couple directly to them, or (with relatively low statistics) states accessible in $B$ decay or $2\gamma$ production.

\begin{table}
\caption {Experimental observations of $X(3872)$.}
\label{tab:X3872}
\begin{center}
\footnotesize
\begin{tabular}{lccccc}
\hline\hline
Expt.\ & Year & Mode & Events & Ref.\\
\hline\hline
Belle  &  2003 & $\pi^+\pi^-J/\psi$ & $35.7\pm6.8$ & \cite{Belle-3872}\\
BaBar & 2004 & $\pi^+\pi^-J/\psi$ &$25.4\pm8.7$ & \cite{BaBar-3872}\\
CDF & 2004 & $\pi^+\pi^-J/\psi$ & $730\pm90$ & \cite{CDF-3872}\\
D\O\ & 2004 & $\pi^+\pi^-J/\psi$ &  $522\pm100$ & \cite{D0-3872}\\
Belle  & 2004 &$\omega (\pi^+\pi^-\pi^0)J/\psi$ & $10.6\pm3.6$& \cite{Belle-3piJ} \\
Belle  & 2005 &$\gamma J/\psi$ & $13.6\pm4.4$ &\cite{Belle-Jpsi-gamma} \\
Belle & 2006&  $D^0{\overline D}{}^{0}\pi^0$ & $23.4\pm5.6$ & \cite{Belle-DDpi}\\
BaBar & 2008 & $\gamma \psi$, $\gamma \psi^\prime$ & $23.0\pm6.4$,$25.4\pm7.3$& \cite{BaBar-psip-gamma}\\
BaBar & 2008&  $D^0{\overline D}{}^{0}\pi^0$ & $33\pm7$ & \cite{BaBar-08b}\\
\hline\hline
\end{tabular}
\end{center}
\end{table}

The final physics example we consider is rare effects in hyperon physics. Two potentially interesting hyperon signals may have been glimpsed in the Fermilab HyperCP experiment, albeit with low statistical significance: evidence for {\em CP} violation   (CPV) in $\Xi^\mp$ decay~\cite{BEACH08} and for flavor-changing neutral currents~\cite{Park-etal} in $\Sigma^+$ decay. While a dedicated experiment to follow up these $<$\,3$\sigma$ effects would be hard to justify, the opportunity for substantial increases in hyperon statistics using the same apparatus that can make the other measurements described here is appealing. As discussed below, other potentially interesting hyperon effects are also within reach of such an experiment and offer a window into new physics  different from those of $K$, $B$, and $D$ studies.

The E835 apparatus did not include a magnet, hence various cross sections needed to assess the performance and reach of a new experiment remain unmeasured. However, they can be estimated with some degree of confidence. We are proposing to assemble, quickly and at modest cost, an ``upgraded E835" apparatus including a magnetic spectrometer. If these cross sections are of the expected magnitude, it should be possible with this apparatus to make the world's best measurements of charm mixing and CPV, as well as of the other effects mentioned above. (To take full advantage of the capabilities of the Fermilab Antiproton Source, a follow-on experiment in a new, dedicated ring \`a la Table~\ref{tab:sens-comp} might  then be designed for even greater sensitivity.)

%\section{Section title}
%\label{sec:1}
%Text with citations \cite{RefB} and \cite{RefJ}.
%\subsection{Subsection title}
%\label{sec:2}
%as required. Don't forget to give each section
%and subsection a unique label (see Sect.~\ref{sec:1}).
%\paragraph{Paragraph headings} Use paragraph headings as needed.
%\begin{equation}
%a^2+b^2=c^2
%\end{equation}
\section{Proposed Antiproton Experiments at Fermilab}

\subsection{Medium-energy $%\mathbf
{\overline p}%\mathbf
 p$-annihilation experiment}
\label{sec:charm}
\begin{figure}
%\vspace{-.1in}
\centerline{%\hspace{-.25in}
\includegraphics[width=\linewidth,bb=120 150 545 460,clip]{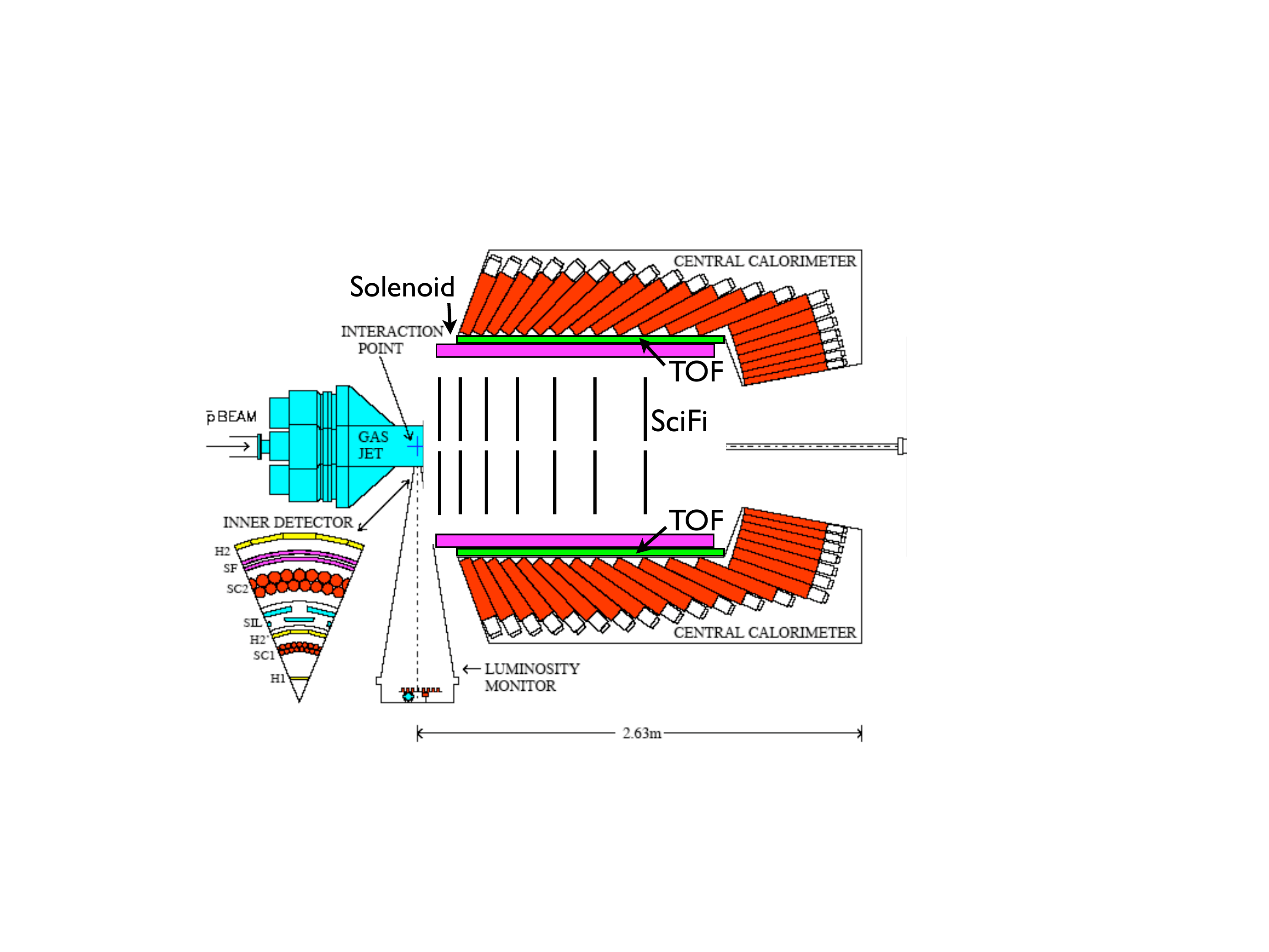}
}
%\vspace{-.5in}
\caption{Sketch of upgraded E835 apparatus as discussed in text: a 1\,T solenoid surrounds fine-pitch scintillating-fiber detectors, 
and is surrounded by precision TOF counters, all within the existing E835 Central Calorimeter.  
A possible return yoke is not shown; if the solenoid is not self-shielding, one would be needed for proper functioning of calorimeter phototubes.}\label{fig:E835-upgrade}
\end{figure}

By adding a small magnetic spectrometer and precision time-of-flight (TOF) counters to the E835 calorimeter as in Fig.\,\ref{fig:E835-upgrade}, plus modern, high-bandwidth triggering and data-acquisition systems, several important topics can be studied. This can be accomplished at modest cost: the solenoid we consider is small  compared to other HEP solenoids, and the very capable scintillating-fiber readout system  from the Fermilab D\O\ experiment~\cite{D0-upgrade-NIM06} should become available once the Tevatron finishes. Cost-effective precision ($\delta t\stackrel{<}{_\sim} 10$\,ps) TOF counters are under development~\cite{Frisch}. We assume $\overline p$$p$ or $\overline p$$N$ luminosity of $2\times10^{32}$\,cm$^{-2}$s$^{-1}$, one order of magnitude beyond that of E835, which can be  accomplished by use of a denser internal target than the E835 hydrogen cluster jet~\cite{E835-NIM}.

% For one-column wide figures use

\paragraph{Charm mixing and  CP violation}

After a more than 20-year search, $D^0$--${\overline D}{}^0$ mixing is now established  at $>$\,10 standard deviations~\cite{DPF2009}, thanks mainly  to the $B$ factories. The level of mixing ($\approx$\,1\%) is consistent with the wide range of standard-model predictions~\cite{Bigi-Uraltsev-Petrov}; however, this does not preclude a significant and potentially detectable contribution from new physics~\cite{Bigi09,Grossman-etal}. Since some new-physics models predict different effects in the charge-2/3 (``up-type") quark sector than in the down-type~\cite{Bigi09,Grossman-etal}, it is important to carry out such  studies not only with $s$ and $b$ hadrons, but with charm mesons as well\,---\,the only up-type system for which  meson mixing can occur.

While  challenging to compute from first principles, recent phenomenological estimates of $\overline{p}p$ annihilation cross sections to open charm near threshold show  $\sigma(\overline{p}p\to D^{*0}{\overline D}{}^0)$ peaking at $\sqrt{s} = 4.2$\,GeV, at about $1\,\mu$b~\cite{Braaten}. (It is interesting to note that the peak of this exclusive cross section fortuitously  occurs at the Antiproton Source design energy.) At ${\cal L}=2\times10^{32}$\,cm$^{-2}$s$^{-1}$, this represents some $4\times10^9$ events produced per year. Since there will  also be $D^{*\pm}D^\mp$, $D^*\overline{D}{}^*$,  $D\overline{D}$,  $D\overline{D}\pi$,...\ events,  the total charm sample will be even larger, and with the use of a target nucleus heavier than hydrogen, the charm-production $A$-dependence~\cite{A-dep} should enhance statistics by a further factor of a few. The total sample could thus substantially exceed the $10^9$ events
now available at the $B$ factories. By localizing the primary interactions to  $\sim$\,10\,$\mu$m along the beam ($z$) direction, a wire target can also allow the $D$-meson decay distance to be cleanly resolved. Medium-energy 
$\overline{p}N$ annihilation may thus be the optimal way to study charm mixing, and to search for possible new-physics contributions via the clean signature~\cite{Petrov,Bigi09} of charm CPV.

We have carried out preliminary simulations of such events with the apparatus of Fig.\,\ref{fig:E835-upgrade}; key parameters of the simulation are given in Table~\ref{tab:params}. In particular we looked at ${\overline p}n \to D^{*-}D^0$, with subsequent decays $D^{*-} \to\pi^-_ s {\overline D}{}^0$, ${\overline D}{}^0 \to K^+\pi^-$, for which the $D^{*-}$ geometric acceptance is about 45\%. To estimate the combinatoric background, we rely on a preliminary analysis of events from the MIPP experiment~\cite{MIPP}, using a 20\,GeV $\overline p$ beam (the lowest energy for which a useful amount of data was available) and scaling the laboratory-frame longitudinal momenta of all secondaries by a factor 0.65 to approximate the effect of running at 8\,GeV.\footnote{The lab-momentum scale factor was determined by comparing the longitudinal-momentum distributions from Monte Carlo simulations of $D^*$ production and decay at 20\,GeV and 8\,GeV beam energy; we note that it is close to the ratio of $\sqrt{s}$ at the two energies. This procedure is conservative in that it neglects the reduction in charged-particle multiplicities and transverse momenta at 8\,GeV compared to 20\,GeV.}  We searched the MIPP data sample for events containing three charged hadrons, two of one sign and one of the other, consistent with being decay products of a $D^{*+} \to\pi^+_ s {D}^0$, ${D}^0 \to K^-\pi^+$ or $D^{*-} \to\pi^-_s {\overline D}{}^0$, ${\overline D}{}^0 \to K^+\pi^-$ decay sequence. We found one such event, corresponding to a continuum cross section of about 1\,$\mu$b before hadron-ID and vertex cuts. 

Given the product branching ratio for the $\pi^\mp K^\pm \pi^\mp$ final state in question and the rarity of charged kaons in 8\,GeV ${\overline p}N$ interactions, we estimate a signal-to-background ratio of about 10-to-1 before vertex cuts. With 150\,$\mu$m resolution in decay-vertex $z$, $>$\,100-to-1 signal-to-background should be possible with efficiency $\stackrel{>}{_\sim}\,$10\%. For example, we could expect to reconstruct $\approx$\,$3\times10^7$ tagged ${D}^0 \to K^-\pi^+$ events per year, to be compared with some $1.2\times10^6$ events in the largest published sample to date~\cite{Belle}, based on 540\,pb$^{-1}$ of data taken at Belle. We also note that the Belle result\,---\,a $D^0\to K\pi$ vs.\ $D^0\to KK/\pi\pi$ lifetime difference of ($1.31\pm0.32\pm0.25$)\%\,---\,has comparable statistical and systematic uncertainties. Thus the precision in a super-$B$ factory may well not improve with increased statistics by as large a factor as  naively expected.

\begin{table}
\begin{center}
\caption{Key detector parameters used in simulations}
\label{tab:params}
\vspace{.1in}
\begin{tabular}{lcc}
\hline\hline
Parameter & value & unit\\\hline
Target ($D$ study):\\
~~ material & Al\\
~~ configuration & wire\\
~~ diameter & 30 & $\mu$m\\
Target ($X$ study):\\
~~ material & H\\
~~ configuration & cluster jet\\
Beam pipe:\\
~~ material & Be\\
~~ diameter & 5 & cm\\
~~ thickness & 350 & $\mu$m\\ 
Solenoid:\\
~~ length & 1.6 & m\\
~~ inner diameter & 90 & cm \\
~~ field & 1 & T\\
SciFi detectors:\\
~~ total thickness per doublet & 360 & $\mu$m\\
~~ fiber pitch & 272 & $\mu$m\\
~~ fiber diameter & 250 & $\mu$m\\
~~ number of stations & 8\\
~~ number of views & 3 \\
~~ number of channels  & $\approx$90,000\\
\hline\hline
\end{tabular}
\end{center}
\end{table}

\paragraph{Hyperon CP violation and rare decays}

The Fermilab HyperCP Experiment~\cite{E871} amassed the world's largest samples of hyperon decays, including $2.5\times10^9$ reconstructed ${}^{^(}\overline{\Xi}{}^{^{\,)\!}}{}^\mp$ decays and $10^{10}$ produced $\Sigma^+$. HyperCP observed unexpected possible signals at the $\stackrel{>}{_\sim}$\,2$\sigma$ level  for new physics in the rare hyperon decay $\Sigma^+\to p\mu^+\mu^-$~\cite{Park-etal} and the  
$\Xi^-\to\Lambda\pi^-$ {\em CP} asymmetry~\cite{BEACH08}: $A_{\Xi\Lambda}=[-6.0\pm 2.1({\rm stat}) \pm 2.0({\rm syst})] \times10^{-4}$ . Since the $\overline{p}p\to\Omega^-{\overline\Omega}{}^+$ threshold lies in the same region as the open-charm threshold, the proposed  experiment  can further test these observations using $\Omega^-\to\Xi^-\mu^+\mu^-$
decays and potential ${}^{^(}\overline{\Omega}{}^{^{\,)\!}}{}^\mp$ CPV (signaled by small $\Omega$--$\overline\Omega$ decay-width differences in ${}^{^(}\overline{\Lambda}{}^{^{\,)\!}}K^\mp$ or ${}^{^(}\overline{\Xi}{}^{^{\,)\!}}{}^0\pi^\mp$ final states)~\cite{OmegaCP}. The HyperCP evidence is suggestive of the range of possible new physics effects. More generally, high-sensitivity hyperon studies  are well motivated irrespective of those ``signals."

\begin{figure}
\vspace{.15in}
\centerline{\hspace{-.075in}
\includegraphics[width=0.44\linewidth]{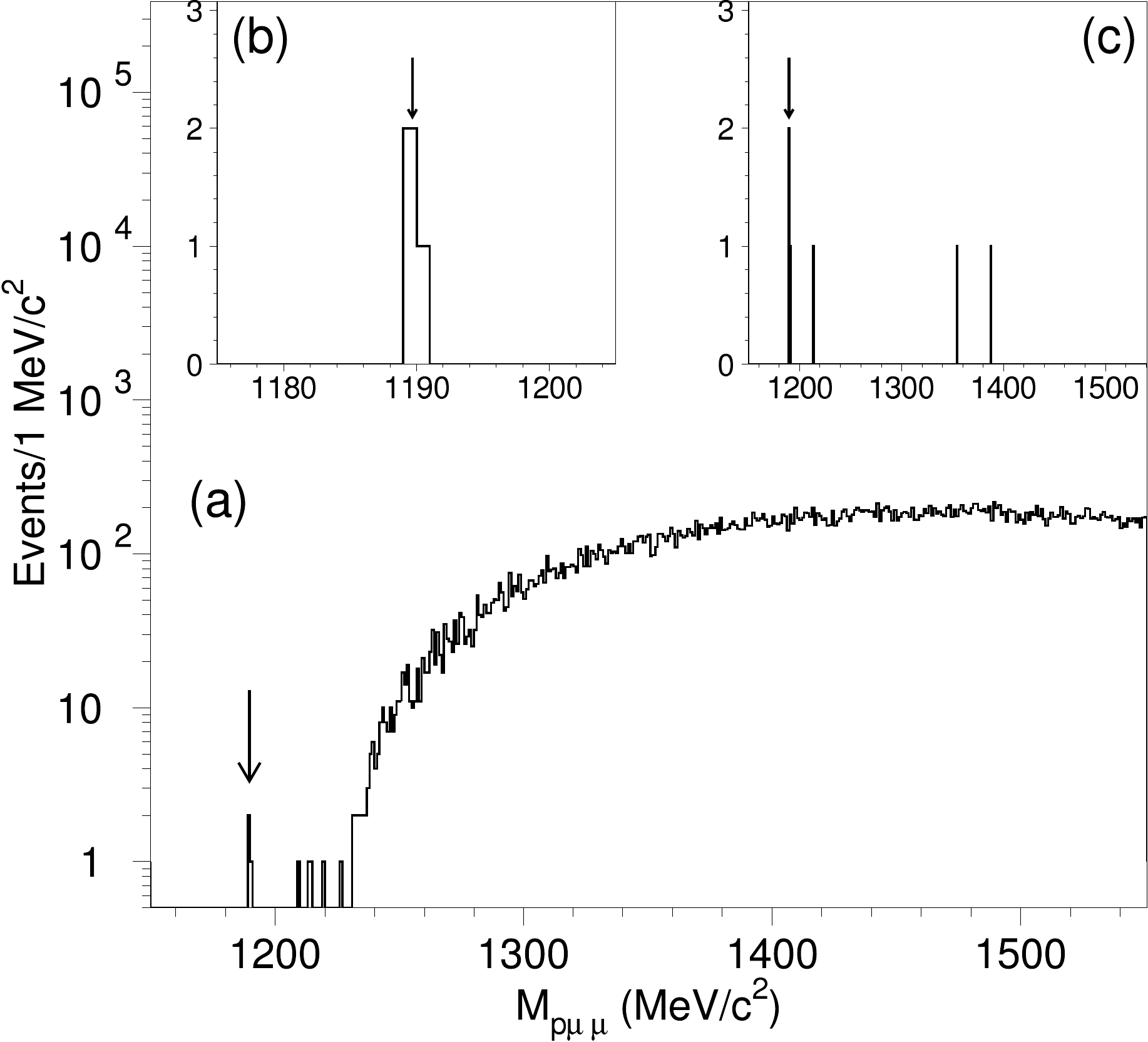}~\includegraphics[width=0.55\linewidth]{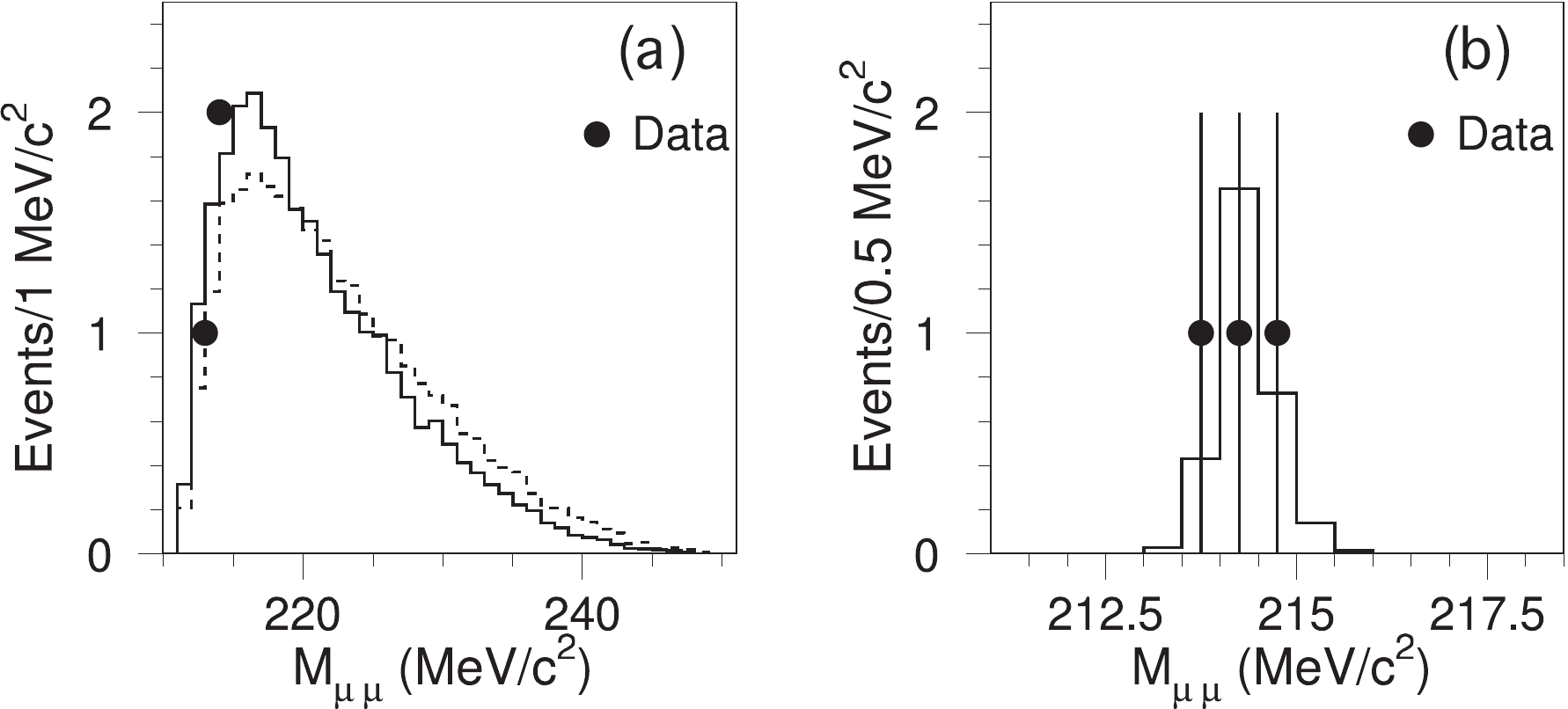}}
\caption{(Left) Mass spectrum for 3-track final states consistent with being single-vertex $p\mu^+\mu^-$  events in HyperCP positive-beam data sample: (a) wide mass range (semilog scale); (b) narrow range around $\Sigma^+$ mass; (c) after application of additional cuts as described in Ref.~\protect\cite{Park-etal}. (Arrows indicate mass of $\Sigma^+$.) (Right) Dimuon mass spectrum of the three HyperCP $\Sigma^+\to p\mu^+\mu^-$ candidate events compared with Monte Carlo spectrum assuming (a) SM virtual-photon form factor (solid) or isotropic decay (dashed), or (b) decay via a narrow resonance $X^0$.
\label{fig:Sigpmumu}}
\end{figure}

While the $\overline p$$p\to \Omega^- {\overline \Omega}{}^+$ cross section has not been measured, by extrapolation from $\overline p$$p\to \Lambda \overline \Lambda$ and $\overline p$$p\to \Xi^- {\overline \Xi}{}^+$ one obtains an estimate just above $\Omega^- {\overline \Omega}{}^+$ threshold of $\approx$\,60\,nb, implying $\sim10^8$ events produced per year. In addition the measured $\approx 1\,$mb cross section for associated production of inclusive hyperons~\cite{Chien-etal} would mean $\sim10^{12}$ events produced per year, which could  directly confront the HyperCP evidence (at $\approx$\,2.4$\sigma$ significance) for a possible new particle of mass 214.3\,MeV/$c^2$ in the three observed $\Sigma^+\to p\mu^+\mu^-$ events (Fig.\,\ref{fig:Sigpmumu}).\footnote{Such a particle, if confirmed, could be evidence for nonminimal SUSY~\protect\cite{Gorbunov}.}   Further in the future, the dedicated $\overline{p}$ storage ring of Table~\ref{tab:sens-comp} could decelerate antiprotons to the $\Lambda\overline{\Lambda}$, $\Sigma^+\overline{\Sigma}{}^-$, and $\Xi^-\overline{\Xi}{}^+$ thresholds, where an experiment at $10^{33}$ luminosity could amass the clean, $>10^{10}$-event samples needed to confirm or refute the HyperCP evidence~\cite{BEACH08} for {\em CP} asymmetry in the $\Xi\Lambda$ decay sequence.

\paragraph{Precision measurements in the charmonium region}

Using the Fermilab Antiproton Source,  experiments E760 and E835 made the world's most precise measurements of charmonium masses and widths~\cite{E760-chi,E835-psi-prime}. This precision ($\stackrel{<}{_\sim}$\,100\,keV) was enabled by the  small energy spread of the stochastically cooled antiproton beam and the absence of Fermi motion and negligible energy loss in the H$_2$ cluster-jet target. 
Although charmonium has by now been extensively studied, a number of questions remain, most notably the nature of the mysterious $X(3872)$ state~\cite{ELQ} and improved measurement  of $h_c$ and $\eta^\prime_c$ parameters~\cite{QWG-Yellow}. The width of the $X$ may well be small compared to 1\,MeV~\cite{Braaten-Stapleton}. The unique precision of the ${\overline p}p$ energy-scan technique is ideally suited to making the precise mass, lineshape, and width measurements needed to test the intriguing hypothesis that the $X(3872)$ is a $D^{*0}\overline{D}{}^0$ molecule~\cite{molecule}.

The production cross section of $X(3872)$ in ${\overline p}p$ annihilation has not been measured, but it has been estimated to be similar in magnitude to that of the $\chi_c$ states~\cite{ditto,Braaten-X-3872}. In E760, the $\chi_{c1}$ and $\chi_{c2}$ were detected in ${\overline p}p \to\chi_c \to\gamma J/\psi$ (branching ratios of 36\% and 20\%, 
respectively~\cite{PDG}) with acceptance times efficiency of $44\pm 2$\%, giving about 500 observed 
events each for an integrated luminosity of 1\,pb$^{-1}$ taken at each resonance; at the mass peak 
1 event per nb$^{-1}$ was observed~\cite{Armstrong-chi_c2}. The lower limit 
${\cal B}[X(3872) \to\pi^+\pi^- J/\psi] > 0.042$ at 
90\% C.L.~\cite{BaBar-BR} implies that in a day at the peak of the $X (3872)$ (8 pb$^{-1} \times$\, [1000 
events/pb$^{-1}$]\,$\times \,0.04/0.36\, \times$\,acceptance-efficiency ratio of final states of $\approx$\,50\%), about 500 
events would be observed. Even if the production cross section is an order of magnitude 
less than those of the $\chi_c$ states, the tens of events per day at the peak will be greater 
than the background observed by E835. By way of comparison, Table~\ref{tab:X3872} shows current 
sample sizes, which are likely to increase by not much more than an order of magnitude 
as these experiments complete during the current decade.\footnote{ 
The ${\overline p}p\to X (3872)$ sensitivity will be competitive even with that of the proposed SuperKEKB~\cite{SuperKEKB}
upgrade, should that project go forward. } (Although CDF and D\O\ could 
amass samples of order $10^4$ $X (3872)$ decays, the large backgrounds in the CDF and D\O\ 
observations, reflected in the uncertainties on the numbers of events listed in Table 2, limit 
their incisiveness.) 

We have concentrated here on one decay mode of the $X (3872)$: $X (3872) \to\pi^+\pi^- J/\psi$.
Large samples will of course also be obtained in other modes as well, increasing the statistics 
and allowing knowledge of $X (3872)$ branching ratios to be improved. Given the uncertainties 
in the cross section and branching ratios, the above may well be an under- or overestimate 
of the ${\overline p}p$ formation and observation rates, perhaps by as much as an order of magnitude. 
Nevertheless, it appears that a new experiment at the Antiproton Accumulator could obtain 
the world's largest clean samples of $X (3872)$, in perhaps as little as a month of running. 
The high statistics, event cleanliness, and unique precision available in the ${\overline p}p$ formation 
technique could enable the world's smallest systematics. Such an experiment could thus 
provide a definitive test of the nature of the $X (3872)$. 

\subsection{Antihydrogen Experiments}
\noindent {\it In-flight {CPT} tests.} Antihydrogen atoms in flight may offer a way around some of the difficulties encountered in the CERN 
trapping experiments. First steps in this direction were taken by Fermilab E862, which observed formation of 
antihydrogen in flight during 1996--7~\cite{Blanford}. Methods to measure the antihydrogen Lamb shift and fine structure (the 
$2s_{1/2}$--$2p_{1/2}$ and $2p_{1/2}$--$2p_{3/2}$ energy differences) were subsequently worked out~\cite{Blanford-Lamb-shift}. Progress towards this goal may be 
compatible with normal Tevatron Collider operations\,---\,a possibility currently under investigation. If the feasibility 
of the approach is borne out by further work, the program could continue into the post-Tevatron era. \\

\noindent {\it Antimatter Gravity Experiment.}
While General Relativity predicts that the gravitational forces on matter and antimatter should be identical, no 
direct experimental test of this prediction has yet been made~\cite{Fischler-etal}. Attempts at a quantized theory of gravity generally 
introduce non-tensor forces, which could cancel for matter-matter and antimatter-antimatter interactions but add for 
matter-antimatter ones. In addition, possible Òfifth forcesÓ or non-$1/r^2$ dependence have been discussed. Such effects 
can be sensitively sought by measuring the gravitational acceleration of antimatter ($\overline g$) in the field of the earth. While 
various such experiments have been discussed for many years, one\,---\,measurement of the gravitational acceleration 
of antihydrogen\,---\,has only recently become feasible and is now proposed both at CERN and at Fermilab~\cite{AEGIS,LoI}.

The principle of the Antimatter Gravity Experiment (AGE), proposed at Fermilab~\cite{LoI}, is to form a beam of slow ($\approx$\,1\,km/s) antihydrogen atoms in 
a Penning trap and pass the beam through an interferometer. The interferometer can employ  material gratings, giving sensitivity to $({\overline g}-g)/g$ at the $10^{-4}$ level, or use laser techniques as pioneered by Chu and Kasevich~\cite{Chu}, with estimated $10^{-9}$ sensitivity. In either approach, a decelerated antiproton beam suitable for trapping is required. With the high antiproton flux available at Fermilab, this can be rather inefficient (say $\sim$\,10$^{-4}$) and still provide enough antiprotons for competitive measurements. Ideas for such deceleration start with the Main Injector, which appears capable of decelerating from 8\,GeV down to $\approx$\,400\,MeV. Below this energy, one possible approach is the ``antiproton refrigerator"~\cite{refrigerator}, employing ``frictional'' cooling. More elaborate, higher-efficiency  solutions, for example a small synchrotron with stochastic cooling, have also been discussed~\cite{LoI}.

\section{Outlook}

With the end of the Tevatron Collider program in sight, new and unique measurements are possible at the Fermilab Antiproton Source~\cite{New-pbar,LEAP08}. If approved, such a program will substantially
 broaden the clientele and appeal of US particle physics. A ``protocollaboration" has been formed and approval is being sought.

%
% For two-column wide figures use
%\begin{figure*}
% Use the relevant command to insert your figure file.
% For example, with the graphicx package use
%  \includegraphics[width=0.75\textwidth]{example.eps}
% figure caption is below the figure
%\caption{Please write your figure caption here}
%\label{fig:2}       % Give a unique label
%\end{figure*}
%
% For tables use
%\begin{table}
% table caption is above the table
%\caption{Please write your table caption here}
%\label{tab:1}       % Give a unique label
% For LaTeX tables use
%\begin{tabular}{lll}
%\hline\noalign{\smallskip}
%first & second & third  \\
%\noalign{\smallskip}\hline\noalign{\smallskip}
%number & number & number \\
%number & number & number \\
%\noalign{\smallskip}\hline
%\end{tabular}
%\end{table}

\section*{Acknowledgements}
%If you'd like to thank anyone, place your comments here
%and remove the percent signs.
% If you have acknowledgments, this puts in the proper section head.
The author thanks his pbar collaborators\footnote{Arizona: A.~Cronin; Riverside: A.~P.~Mills, Jr.; Cassino: G.~M.~Piacentino; Duke: T.~J.~Phillips; FNAL: G.~Apollinari, D.~R.~Broemmelsiek, B.~C.~Brown, C.~N.~Brown, D.~C.~Christian, P.~Derwent, M.~Fischler, K.~Gollwitzer, A.~Hahn, V.~Papadimitriou,  G.~Stancari, M.~Stancari, R.~Stefanski, J.~Volk, S.~Werkema, H.~B.~White, G.~P.~Yeh; Ferrara: W.~Baldini; Hbar Tech.: J.~R.~Babcock, S.~D.~Howe, G.~P.~Jackson, J.~M.~Zlotnicki; IIT: D.~M.~Kaplan, T.~J.~Roberts, H.~A.~Rubin, Y.~Torun, C.~G.~White; KSU: G.~A.~Horton-Smith, B. Ratra; KyungPook: H.~K.~Park; Luther Coll.: T.~K.~Pedlar; Michigan: H.~R.~Gustafson, M.~Longo, D.~Rajaram; Northwestern: J.~Rosen; Notre Dame: M. Wayne; SMU: T.~Coan; SXU: A. Chakravorty; Virginia: E.~C.~Dukes; Wayne State: G.~Bonvicini.
} for their support and encouragement, and E. Braaten, E. Eichten, 
and C. Quigg for useful conversations. 
Work supported by Department of Energy grant DE-FG02-94ER40840.
%\end{acknowledgements}

% BibTeX users please use one of
%\bibliographystyle{spbasic}      % basic style, author-year citations
%\bibliographystyle{spmpsci}      % mathematics and physical sciences
%\bibliographystyle{spphys}       % APS-like style for physics
%\bibliography{}   % name your BibTeX data base

% Non-BibTeX users please use

\end{document}